\documentclass{article}

\usepackage[final, nonatbib]{neurips_2021}

\usepackage[utf8]{inputenc} 
\usepackage[T1]{fontenc}    
\usepackage{hyperref}       
\usepackage{url}            
\usepackage{booktabs}       
\usepackage{amsfonts}       
\usepackage{nicefrac}       
\usepackage{microtype}      
\usepackage{xcolor}         
\usepackage{cite}
\usepackage{subcaption} 
\usepackage{graphicx}
\usepackage{multirow}

\newcommand{\ie}{\textit{i.e.}, \ }
\newcommand{\eg}{\textit{e.g.}, \ }

\newcommand{\Albatross}{Aquarius\ }
\newcommand{\Albatrosss}{Aquarius}

\title{Towards Intelligent Load Balancing in Data Centers}

\author{%
  Zhiyuan Yao$^{1,2}$, Yoann Desmouceaux$^{2}$, Mark Townsley$^{2}$, Thomas Clausen$^{1}$\\\\
  $^{1}$\'Ecole Polytechnique\\
  $^{2}$Cisco Systems
}

\begin{document}

\maketitle

\begin{abstract}
  Network load balancers are important components in data centers to provide scalable services.
  Workload distribution algorithms are based on heuristics, \eg Equal-Cost Multi-Path (ECMP), Weighted-Cost Multi-Path (WCMP) or naive machine learning (ML) algorithms, \eg ridge regression.
  Advanced ML-based approaches help achieve performance gain in different networking and system problems.
  However, it is challenging to apply ML algorithms on networking problems in real-life systems.
  It requires domain knowledge to collect features from low-latency, high-throughput, and scalable networking systems, which are dynamic and heterogenous.
  This paper proposes \Albatross to bridge the gap between ML and networking systems and demonstrates its usage in the context of network load balancers.
  This paper demonstrates its ability of conducting both offline data analysis and online model deployment in realistic systems.
  The results show that the ML model trained and deployed using \Albatross improves load balancing performance yet they also reveals more challenges to be resolved to apply ML for networking systems.
\end{abstract}

\section{Introduction}
\label{sec:intro}

In data centers, applictions are replicated on multiple instances running \eg in containers or virtual machines (VMs) to provide scalable services~\cite{dragoni2017microservices}.
One of the main components in such data centers for optimal resource utilization is \emph{network} (load balancers), whose role is to distribute network traffic \textit{fairly} among application instances.
As ML-based approaches achieve performance gains in different networking problems~\cite{usama2017unsupervised, xie2018survey}, this paper investigates whether ML helps improve network load balancing performance.

The challenges of applying ML on network load balancing problem, especially in real-world systems, are $3$-fold.
First, feature collections require domain knowledge.
Unlike task schedulers~\cite{auto2018sigcomm} or application-level load balancers~\cite{yoda}, network load balancers have limited observations and are not aware of task size and duration before distributing workloads.
They can only extract features from packet headers below the transport layer.
Second, networking systems favor low-latency, high-throughput and scalability.
Adding ML algorithms in the system incurs additional computational overhead for collecting and processing features, making predictions, and online training, which degrades data plane performance and scalability~\cite{taurus2020}.
Third, networking environments are dynamic and heterogenous~\cite{kumar2020fast, fu2021use}.
Asynchronous closed-loop design is required to bring ML algorithms online so that the models can be adapted over time without blocking the system.

This paper proposes \Albatross to bridge the different requirements for networking system and ML.
\Albatross is an asynchronous and scalable data collection and exploitation mechanism that enables ML-based decisions based on fine-grained observations.
This paper implements \Albatross in Vector Packet Processing (VPP)~\cite{vpp}, which is easy to deploy in real-world systems.
Using \Albatrosss, the potential benefits and challenges of ML for network load balancing problem are investigated.

\section{Related Work}
\label{sec:related}

ML techniques (\eg graph neural networks~\cite{decima2018}, and convolutional neural networks~\cite{naseer2018enhanced}) help optimize and classify network traffic in data centers.
However, applying these techniques alongside the data plane on the fly is computationally intractable~\cite{ahmed2016survey, taurus2020}.
In~\cite{mvfst-rl}, it is shown that asynchronous design helps achieve performance gain without degrading networking performance on emulators.
This paper implements \Albatross on a platform compatible to commodity CPUs so that it can be deployed in real-world system.

In~\cite{fu2021use}, the challenges of applying ML algorithms on networking systems are studied using system configurations as features.
In the context of network load balancing\cite{maglev}, ridge regression~\cite{lbas-2020} is used to improve workload distribution fairness using actively probed server load information (CPU and memory usage) as features.
With \Albatrosss, the same problem can be investigated using a wide range of runtime networking features extracted from packet headers, which makes load balancers no longer necessary to maintain the active probing channel with all servers.

\section{Overview}
\label{sec:overview}

\begin{figure}[t]
  \centering
  \begin{minipage}{.31\textwidth}
    \centering
    \includegraphics[height=2.35in]{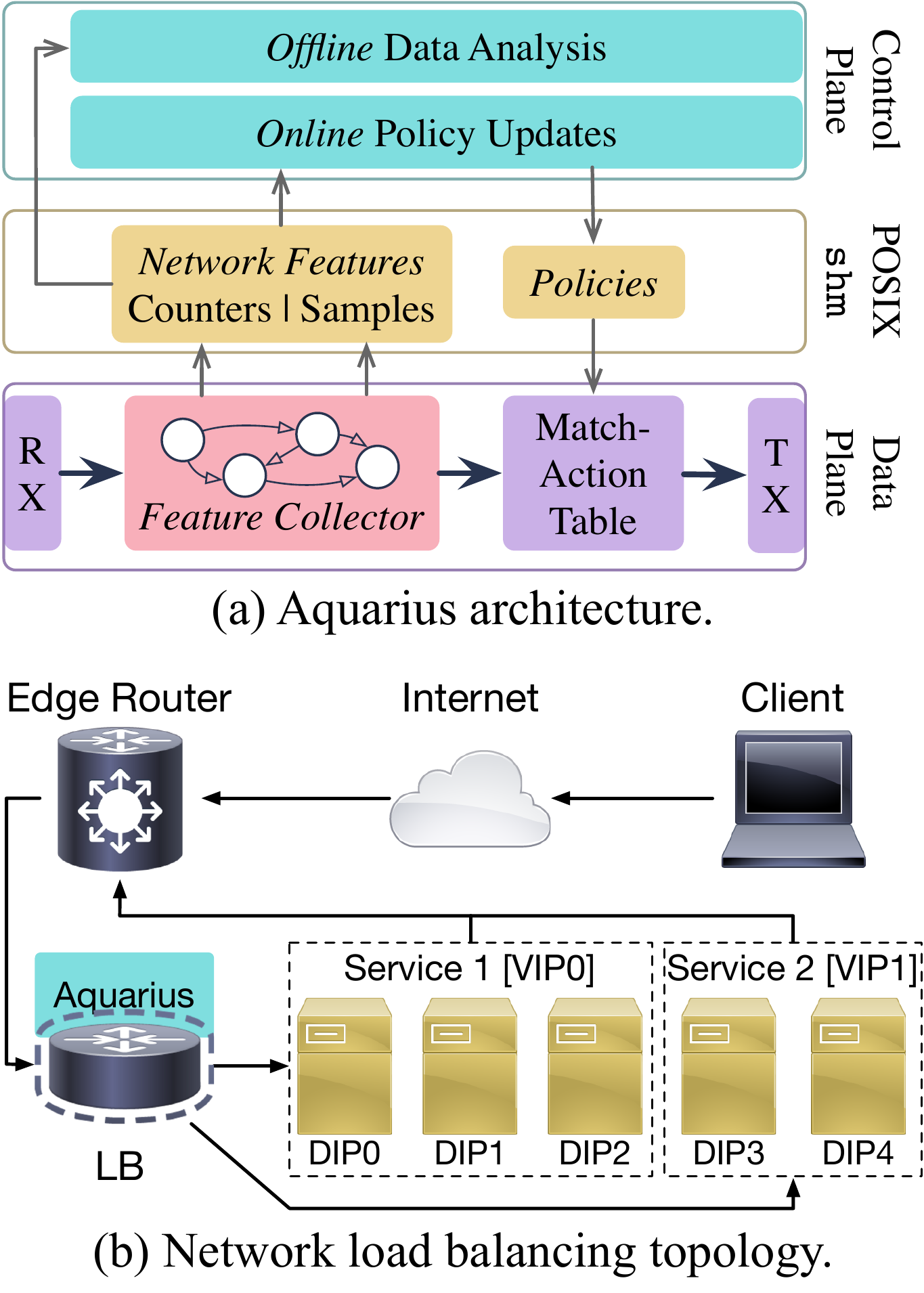}
    \caption{Overview.}
    \label{fig:overview}
  \end{minipage}%
  \hspace{.15in}
  \begin{minipage}{.65\textwidth}
    \centering
    \includegraphics[height=2.35in]{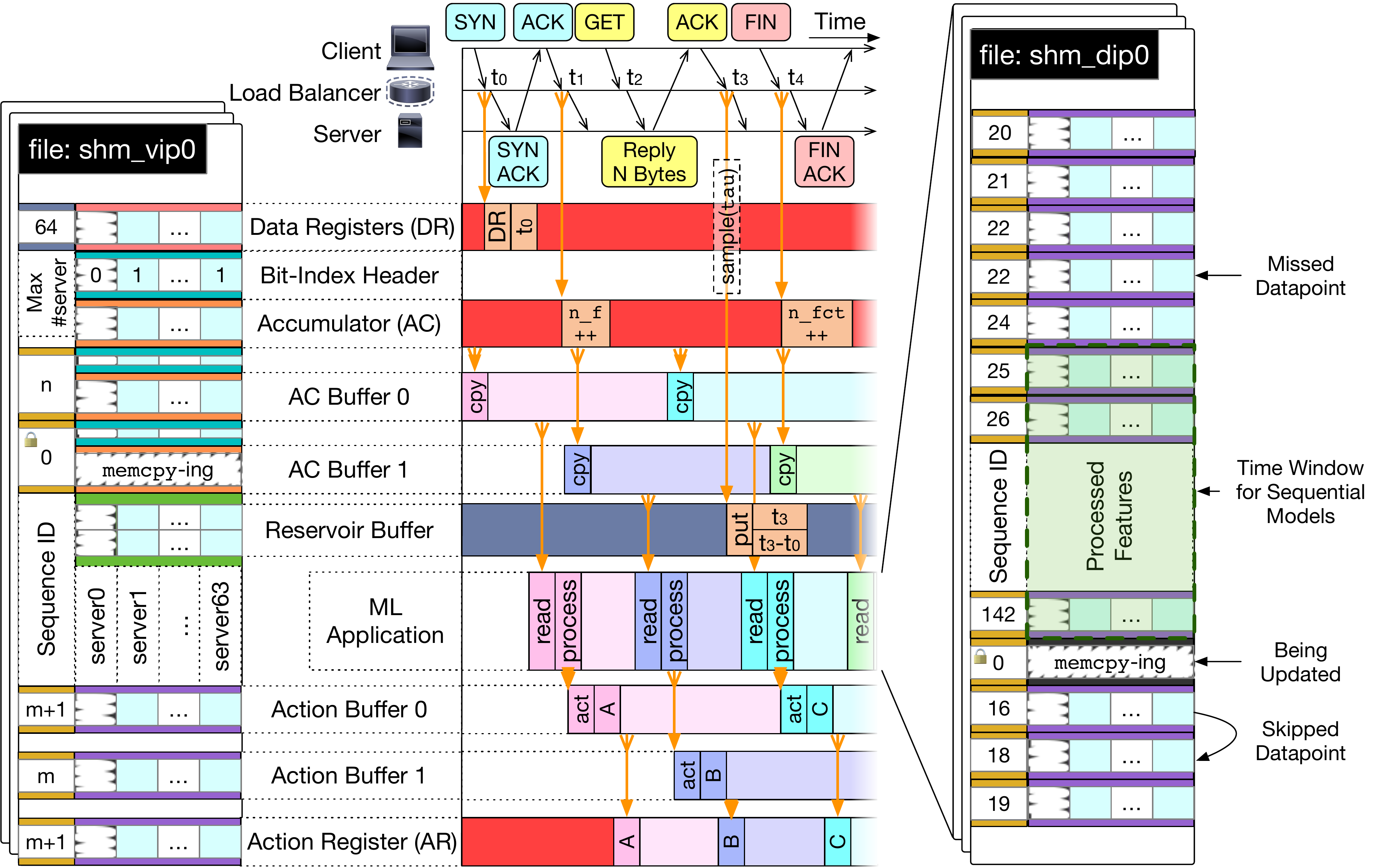}
    \caption{\Albatross \texttt{shm} layout and data flow pipeline.}
    \label{fig:design}
  \end{minipage}
  \vskip -.15in
\end{figure}

\Albatross has a $3$-layer architecture (figure~\ref{fig:overview}a).
It extracts network features from the data plane (parser layer) and makes the features available via shared memory (partitioner layer) for the control plane (processor layer).

In the context of network load balancing problem, \Albatross is deployed on load balancers.
As depicted in figure~\ref{fig:overview}b, cloud services provided in data centers are identified by virtual IPs (VIPs).
Each VIP corresponds to a cluster of virtualized application servers, each identified by a unique direct IP (DIP).
Within each VIP, \Albatross needs to track the state of each server to distinguish the overloaded or malfunctioning ones and make more-informed load balancing decisions.

\section{Design}
\label{sec:design}

In order to apply ML in an asynchronous close-loop load balancing framework with high scalability and low latency, communication between the load balancer data plane and the ML application is implemented via POSIX shared memory (\texttt{shm}).
The design of \Albatross allows features to be extracted from the data plane and conveyed to the ML application, and allows data-driven decisions generated by the ML application to be updated asynchronously on the load balancer.

The pipeline of the data flow over the lifetime of a TCP connection is depicted in figure \ref{fig:design}.
On receipt of different networking packets, networking features are gathered as counters or samples.
To avoid I/O conflicts, sampled features are collected using reservoir sampling over the latest time window and counters are collected atomically and made available to the data processing agent using multi-buffering.
The bit-index binary header helps efficiently identify active application servers.
Gathered features are organized by the packets' corresponding VIP and DIP in \texttt{shm} files identified by VIP (\eg \texttt{shm\_vip0}).
With no disruption in the data plane, these features are fetched by ML application periodically to \texttt{shm} files identified by DIP (\eg \texttt{shm\_dip0}), which serve as a database for the ML application.
Only the features with the highest sequence ID are fetched and sequence ID $0$ is used as a writing ``lock''.
Using the same multi-buffering scheme, action buffes and registers allow to effectuate policies generated by the ML application.

This design is asynchronous and has no blocking steps.
This design also favors the discrete arrivals of networking packets, and allows to gather $21$ networking features\footnote{The whole list of networking features are listed in figure \ref{fig:app-feature} in appendix \ref{app:feature}.}.
This design separates gathered networking features by VIP and DIP, and allows to aggregate the features at different levels and make predictions for different purposes.
Updating (adding or removing) services (VIPs) and their associated servers (DIPs) can be achieved by managing different \texttt{shm} files in a scalable way using this design, incurring no disruption on data planes.

\section{Experiments}
\label{sec:experiments}

Using the same topology as in figure \ref{fig:overview}b, $3$ different network traces\footnote{Wiki, Poisson \texttt{for}-loop and file traces. See appendix \ref{app:testbed} for details.} are applied as network traffic over $2$ groups of ($2$-CPU and $4$-CPU) servers with different processing capacities.
Throughout the set of experiments, network features ($8$ counters and $13$ sampled features) are collected as input data for ML models to predict $3$ ground truth values, \ie number of provisioned CPUs (\texttt{\#cpu}), CPU usage, and number of busy worker threads (\texttt{\#thread}).
Each sampled feature channel is reduced to $5$ scalars, \ie average, $90$th-percentile, standard deviation, exponential moving average (\texttt{decay}) of average and $90$th-percentile.
This section illustrates both offline (section \ref{sec:experiments-offline}) and online (section \ref{sec:experiments-online}) application of \Albatross for developing an ML-based load balancer.

\subsection{Offline ML Applications}
\label{sec:experiments-offline}

An ECMP load balancer is implemented with \Albatrosss.
Features and ground truth values are collected every $50$ms along with the different types of input network traffic with different traffic rate.

\textbf{Feature process pipeline}: Collected dataset is preprocessed and converted to have zero mean and unit standard deviation.
They are subtracted by the mean and divided by the standard diviation across the entire training set.
Outlier data-points (any feature or ground truth value beyond $99$th-percentile) are dropped.

\begin{figure}[t]
	\begin{center}
		\centerline{\includegraphics[width=\columnwidth]{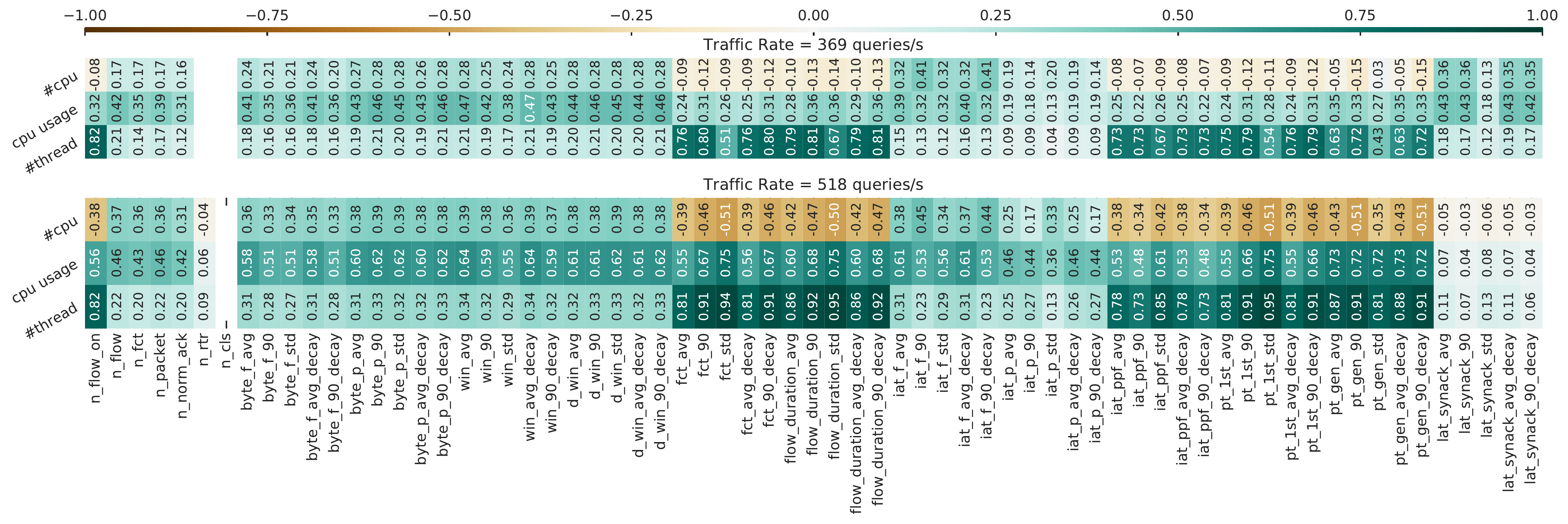}}
		\caption{Feature correlation obtained from Wikipedia replay traces applied on a $20$-CPU Apache server clusters where $2$ groups of servers have different provisioned capacities ($2$-CPU and $4$-CPU).}
		\label{fig:offline-corr}
	\end{center}
	\vskip -0.25in
\end{figure}

\begin{figure}[t]
  \centering
  \begin{minipage}{.5\textwidth}
    \centering
    \includegraphics[height=1.4in]{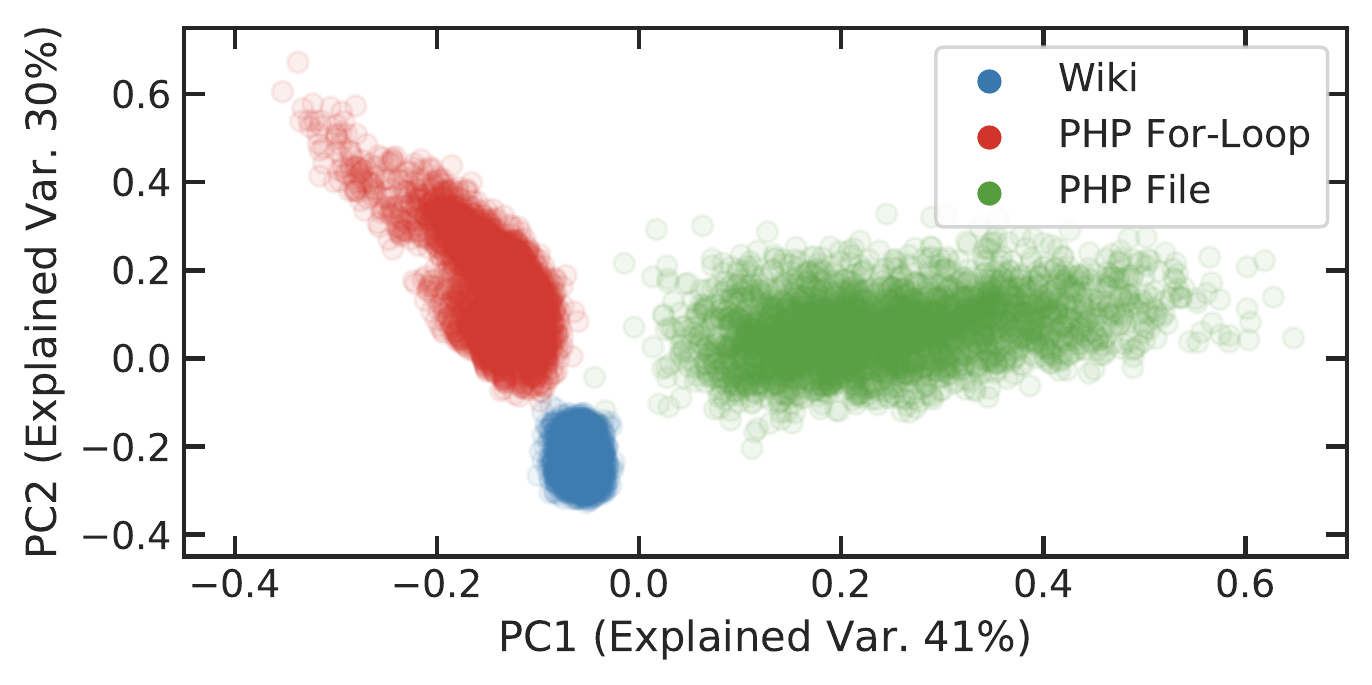}
    \caption{Network traces clustering with PCA.}
    \label{fig:offline-pca}
  \end{minipage}%
  \begin{minipage}{.47\textwidth}
    \centering
    \includegraphics[height=1.4in]{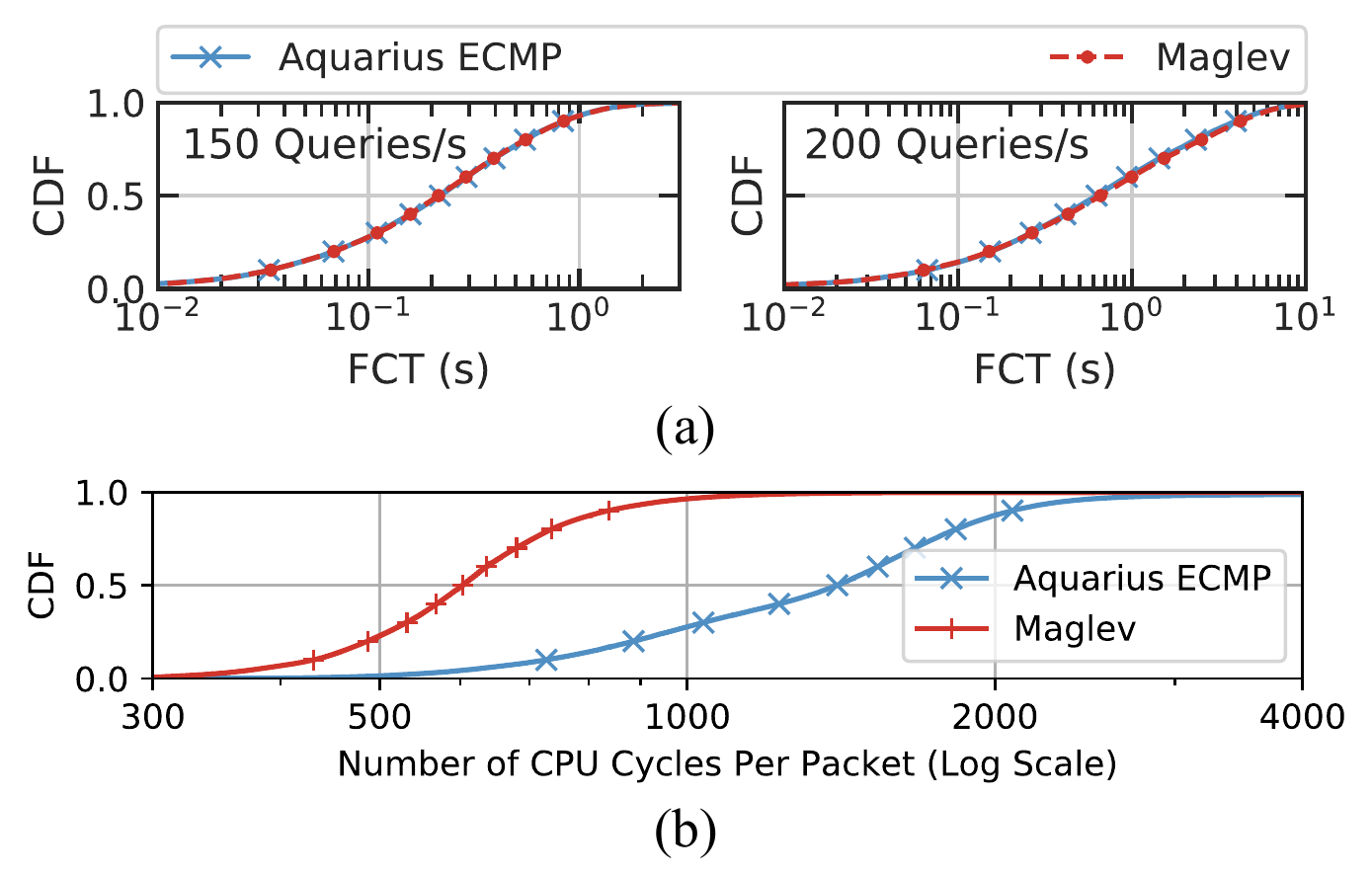}
    \caption{\Albatross overhead analysis.}
    \label{fig:offline-overhead}
  \end{minipage}
  \vskip -.3in
\end{figure}

\textbf{Data analysis}: Correlation between networking features and the $3$ ground truth values using Wiki trace are plotted in figrue \ref{fig:offline-corr}.
It can be observed that under higher traffic rate, the flow completion time (\texttt{fct}) and flow duration have higher positive correlation with the actual server load states (CPU usage and \texttt{\#thread}) and negative correlation with the provisioned processing capacities.
This makes sense since under heavy workloads, servers processing speed decreases and more powerful servers finish tasks faster.
Conducting principal component analysis (PCA) on the collected networking features using $3$ types of traffic gives clustering results as depicted in figure \ref{fig:offline-pca}.
Projected on two principal components (PCs), which accounts for $41\%$ and $30\%$ of the overall variability, $3$ clusters can be clearly observed.
This is a promising result for potential ML-based traffic classifiers, which distinguish traffic types and allocate different computational resources to meet corresponding requirements of quality of service (QoS).

\textbf{Overhead analysis}: The performance of \Albatross is compared with state-of-the-art Maglev load balancer~\cite{maglev}.
As depicted in figure \ref{fig:offline-overhead}a, \Albatross does not induce notable degradation of QoS.
\Albatross introduces an additional $1$k per-packet processing CPU cycles ($0.385\mu$s on a $2.6$GHz CPU) on average (figure \ref{fig:offline-overhead}b), which is trivial comparing with the typical round trip time (higher than $200\mu$s) between network equipments~\cite{guo2015pingmesh}.

\textbf{Training}: $8$ ML models are trained to predict \texttt{\#thread} as server load estimators to make load-aware load balancing decisions.
To adapt the dataset for sequential models, the sequence length is $64$ and stride is $32$, which give $160$k datapoints in total.
These sequential datapoints are randomly splited $80:20$ into training and testing datasets.
Tensorflow~\cite{tensorflow2015-whitepaper} is used for model training.
The hyperparameters for different models are described in appendix \ref{app:model}.

\textbf{Results}: As shown in table \ref{tab:ml-score}, recurrent models have better performance in general (\ie LSTM and RNN).
Applying convolutional layers helps reduce inference delay (\ie 1DConv-GRU1).
More complicated models do not necessarily improve model performance (\ie Wavenet models).

\begin{table}[t]
	\caption{Accumulated score board for different models and different tasks executed w/ 1 CPU core.}
	\label{tab:ml-score}
  \centering
  \resizebox{\textwidth}{!}{ 
    \begin{tabular}{ccccccccc}
      \toprule
      Task                             & Metrics                                     & Dense1 & RNN2 & LSTM2 & GRU2  & \begin{tabular}[c]{@{}c@{}}1DConv-\\ GRU1\end{tabular} & \begin{tabular}[c]{@{}c@{}}Wavenet-\\ GRU1\end{tabular} & \begin{tabular}[c]{@{}c@{}}Wavenet-\\ Reconst.\end{tabular} \\ \midrule
      \multirow{7}{*}{\rotatebox[origin=c]{90}{\hspace{3em}Wiki}}     & MSE                                                              & $253.203$         & $2.557$         & $\mathbf{1.553}$ & $1.660$          & $1.878$                                                           & $1.923$                                                            & $2.421$                                                                \\
      & RMSE                                                             & $15.912$          & $1.599$         & $\mathbf{1.245}$ & $1.288$          & $1.371$                                                           & $1.387$                                                            & $1.556$                                                                \\ 
      & MAE                                                              & $1.804$           & $1.099$         & $\mathbf{0.916}$ & $0.931$          & $0.988$                                                           & $0.996$                                                            & $1.117$                                                                \\ 
      & Delay (ms)  & $50.5\pm1.0$            & $52.1\pm1.7$          & $53.4\pm3.0$           & $53.7\pm1.3$           & $\mathbf{44.8\pm2.8}$                                                   & $54.9\pm0.8$                                                             & $54.6\pm0.5$                                                                 \\ \midrule
      \multirow{7}{*}{\rotatebox[origin=c]{90}{\hspace{3.2em}Poisson}} & MSE                                                              & $1520.888$        & $2.804$         & $0.801$          & $\mathbf{0.774}$ & $0.874$                                                           & $0.965$                                                            & $0.946$                                                                \\
      & RMSE                                                             & $38.999$          & $1.675$         & $0.895$          & $\mathbf{0.880}$ & $0.935$                                                           & $0.982$                                                            & $0.973$                                                                \\
      & MAE                                                              & $3.176$           & $1.162$         & $0.602$          & $\mathbf{0.600}$ & $0.635$                                                           & $0.648$                                                            & $0.649$                                                                \\
      & Delay (ms)  & $\mathbf{55.4\pm0.7}$   & $91.4\pm6.4$          & $69.3\pm2.2$           & $70.9\pm1.3$           & $61.5\pm2.5$                                                            & $65.6\pm1.7$                                                             & $70.2\pm0.4$  \\ \bottomrule
    \end{tabular}
  }
  \vskip -.2in
\end{table}

\subsection{Online ML Applications}
\label{sec:experiments-online}

\begin{figure}[t]
	\centering
	\begin{subfigure}{.31\columnwidth}
		\centerline{\includegraphics[height=.45\columnwidth]{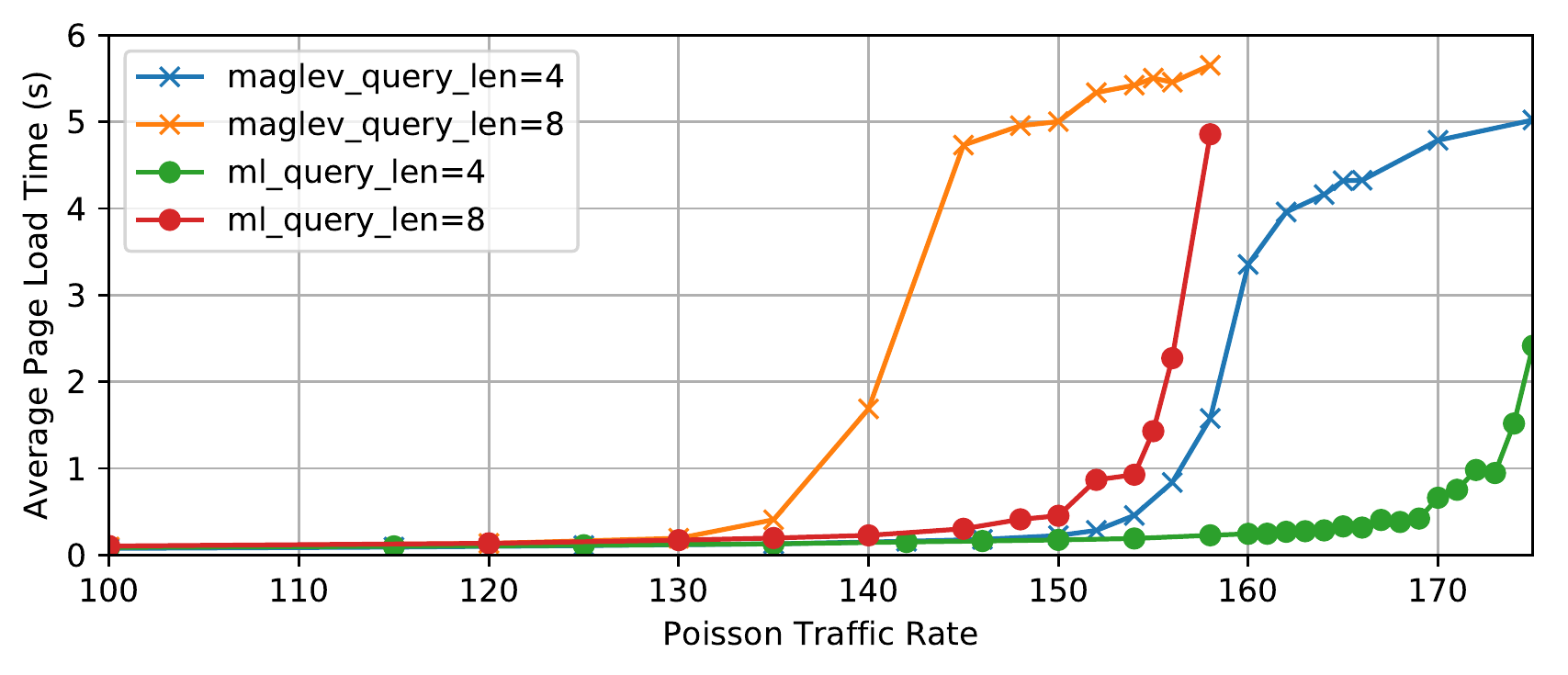}}
		\caption{Page load time}
		\label{fig:online-plt-avg}
	\end{subfigure}%
	\hspace{.1in}
	\begin{subfigure}{.31\columnwidth}
		\centerline{\includegraphics[height=.45\columnwidth]{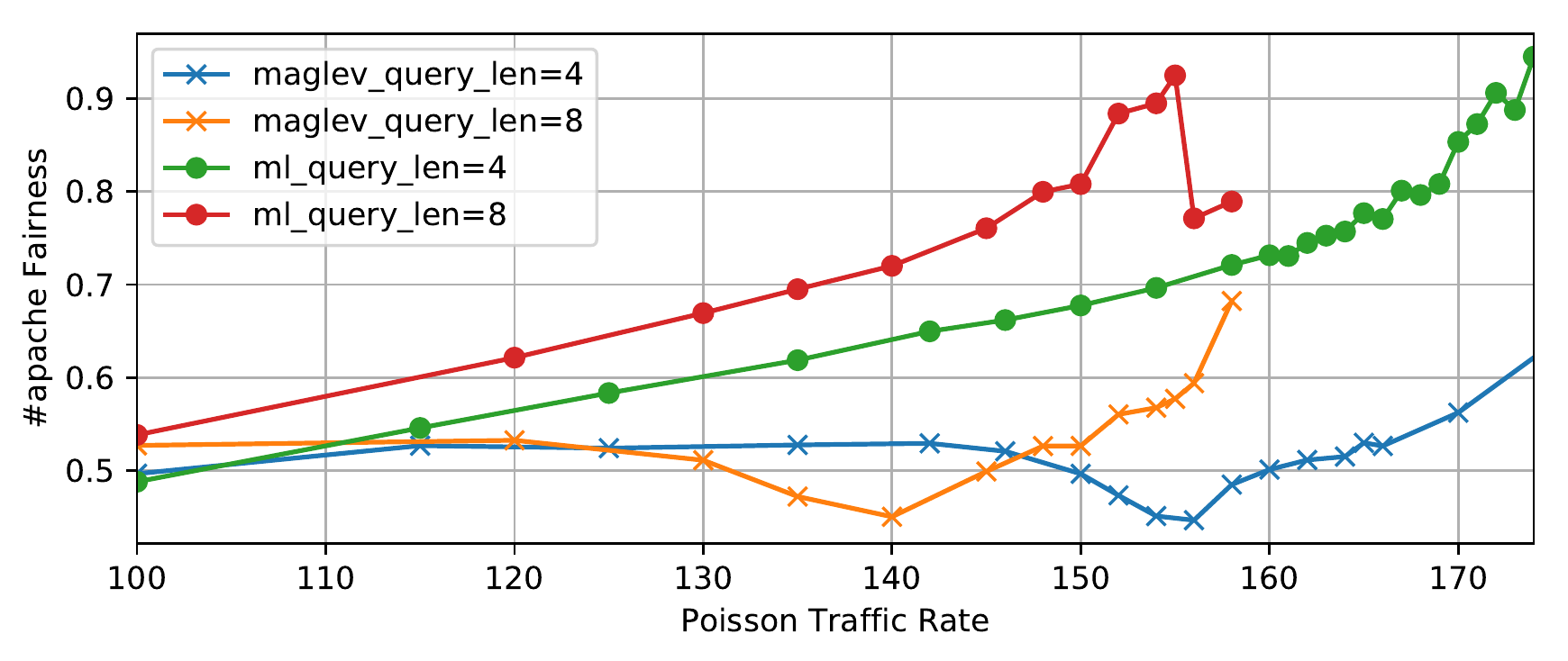}}
		\caption{Jain's fairness of \texttt{\#thread}.}
		\label{fig:online-apache-fair}
	\end{subfigure}%
	\hspace{.1in}
	\begin{subfigure}{.31\columnwidth}
		\centerline{\includegraphics[height=.45\columnwidth]{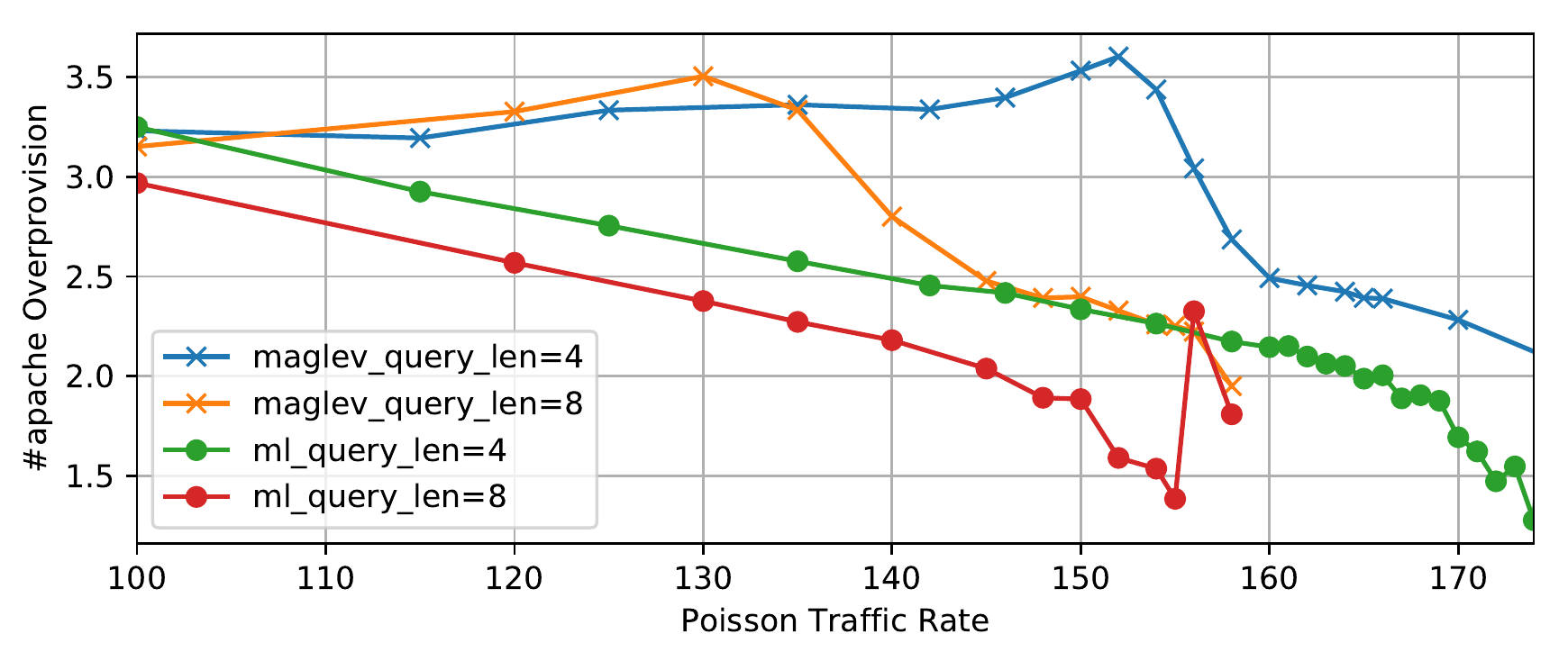}}
		\caption{\texttt{\#thread} overprovision.}
		\label{fig:online-apache-over}
	\end{subfigure}%
	\vspace{.1in}
	\begin{subfigure}{.31\columnwidth}
		\centerline{\includegraphics[height=.45\columnwidth]{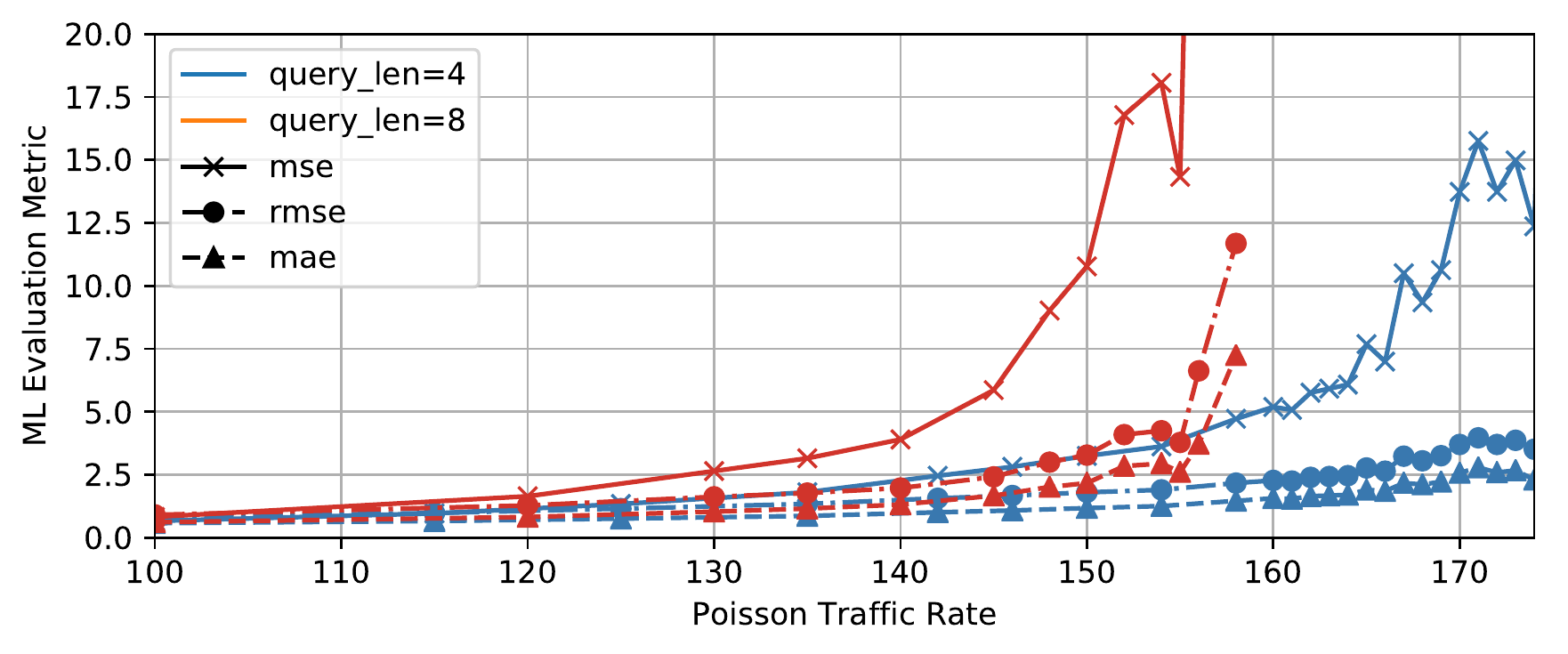}}
		\caption{ML score.}
		\label{fig:online-ml-score}
	\end{subfigure}%
	\hspace{.1in}
	\begin{subfigure}{.31\columnwidth}
		\centerline{\includegraphics[height=.45\columnwidth]{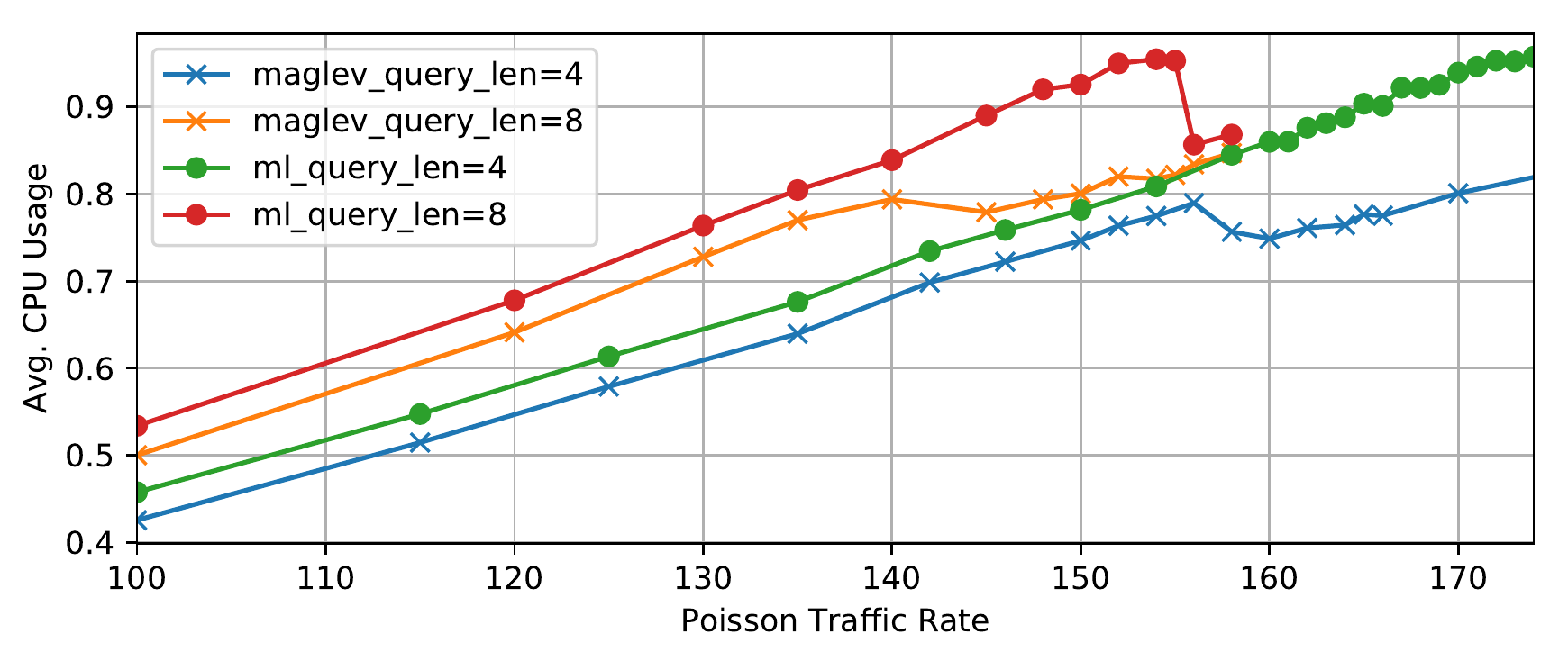}}
		\caption{Avg. CPU.}
		\label{fig:online-cpu-avg}
	\end{subfigure}%
	\hspace{.1in}
	\begin{subfigure}{.31\columnwidth}
		\centerline{\includegraphics[height=.45\columnwidth]{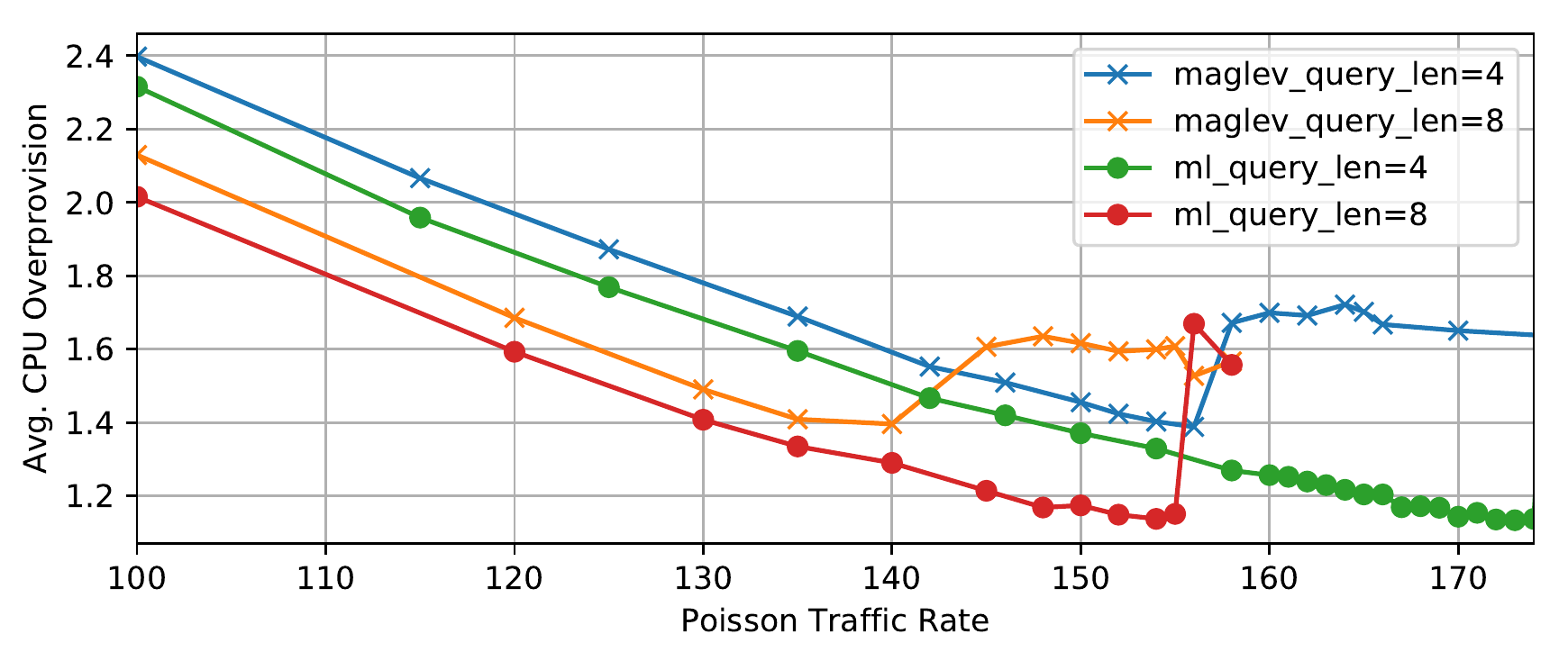}}
		\caption{CPU overprovision.}
		\label{fig:online-cpu-over}
	\end{subfigure}%
	\caption{Online load balancing with trained LSTM2 model on Poisson \texttt{for}-loop trace.}
	\label{fig:online}
  \vskip -.2in
\end{figure}

\textbf{Online performance}: \Albatross enables open and closed-loop control.
Based on the results from previous section, LSTM2 model is brought online to make load balancing decisions (changing the server weights and assigning more tasks to servers with higher weights) based on latest observation every $250$ms.
Two Poisson \texttt{for}-loop traces are applied as input traffic.
The average completion times for each query are $140$ms for \texttt{query\_len=4} and $160$ms \texttt{query\_len=4}).
The load balancing performance is compared with Maglev in figure \ref{fig:online} across a wide range of traffic rates.
It is shown that with trained ML models, load balancers allow the same server cluster to serve heavier workloads with reduced page load time (\texttt{fct}), by optimizing resource utilization (improving workload distribution fairness and reducing overprovision factor).

\begin{figure}[t]
	\centering
  \begin{subfigure}{.45\columnwidth}
    \centering
    \includegraphics[width=\columnwidth]{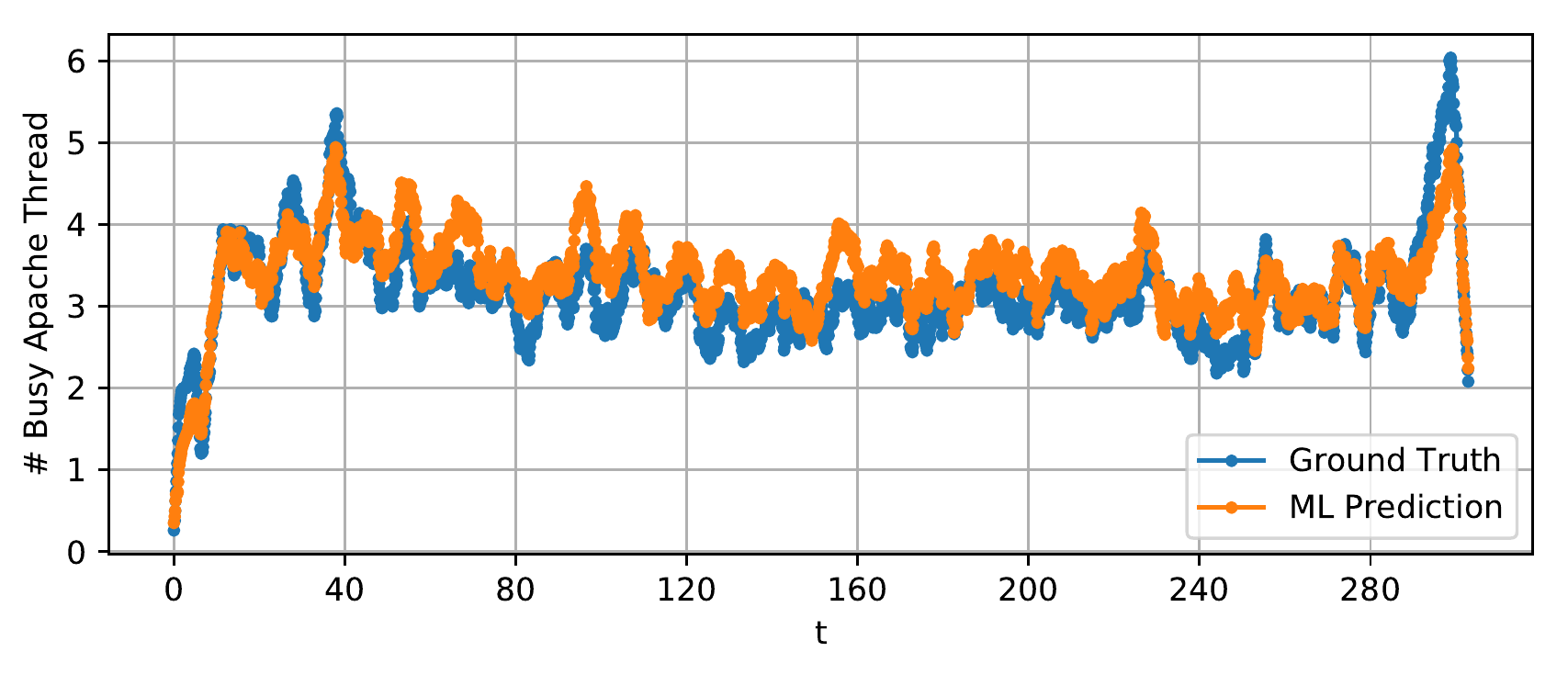}
    \caption{Poisson \texttt{for}-loop traffic.}
    \label{fig:general-architecture}
  \end{subfigure}%
  \hspace{.2in}
  \begin{subfigure}{.45\columnwidth}
    \centering
    \includegraphics[width=\columnwidth]{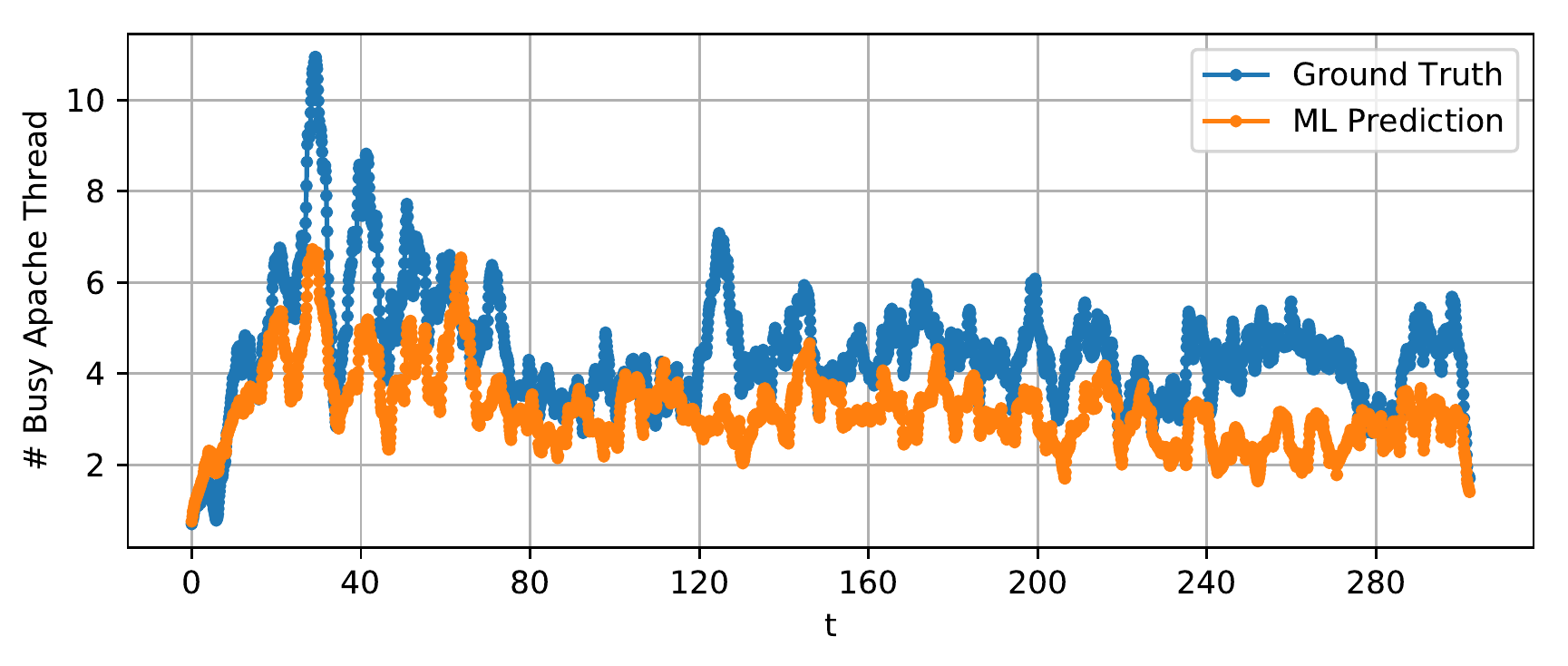}
    \caption{Wiki traffic.}
    \label{fig:general-lb}
  \end{subfigure}
	\caption{Predicting $2$ types of network traffic with LSTM2 model trained only with \texttt{for}-loop trace.}
  \vskip -.2in
	\label{fig:general}
\end{figure}

\textbf{Generality}: To study whether the trained ML models is able to generalize, model LSTM2 is trained only using Poisson \texttt{for}-loop traffic and brought online to work with both \texttt{for}-loop and Wiki traces.
Figure \ref{fig:general} shows that the trained model generalizes poorly if the applied traffic is not seen by the model before, which is consistent to~\cite{fu2021use}.

\section{Conclusion and Future Work}
\label{sec:conclusion}

ML algorithms show promising results on different problems yet it is challenging to apply on realistic networking problems and in real-life systems.
This paper proposes \Albatross to bridge ML and distributed networking systems and takes a preliminary step to integrate ML approaches in networking field.
It allows to gather a wide range of networking features and feed them to offline data analysis pipelines or online ML models to make decisions on the fly.
The results demonstrates the potential of \Albatross to conduct feature engineering, traffic classification, model selection, and online model deployment.
The models applied in this paper shows the ability to learn and infer server load states with networking features.
It also shows that networking problems are dynamic and heterogenous, thus it is challenging to train a model that generalizes well.
Reinforcement learning will be studied in future work to improve model generality in the interactive real-world system.
This work has several limitations.
ML models and their hyperparameters are not sufficiently explored.
The asynchronous decisions are delayed and impacts of delayed decisions along with action updating frequencies are not fully investigated.



\bibliographystyle{unsrt}
\bibliography{reference}

\newpage
\appendix

\section{Feature Collection}
\label{app:feature}

\begin{figure}[t]
  \centering
  \begin{minipage}{.45\textwidth}
    \centering
    \includegraphics[height=3in]{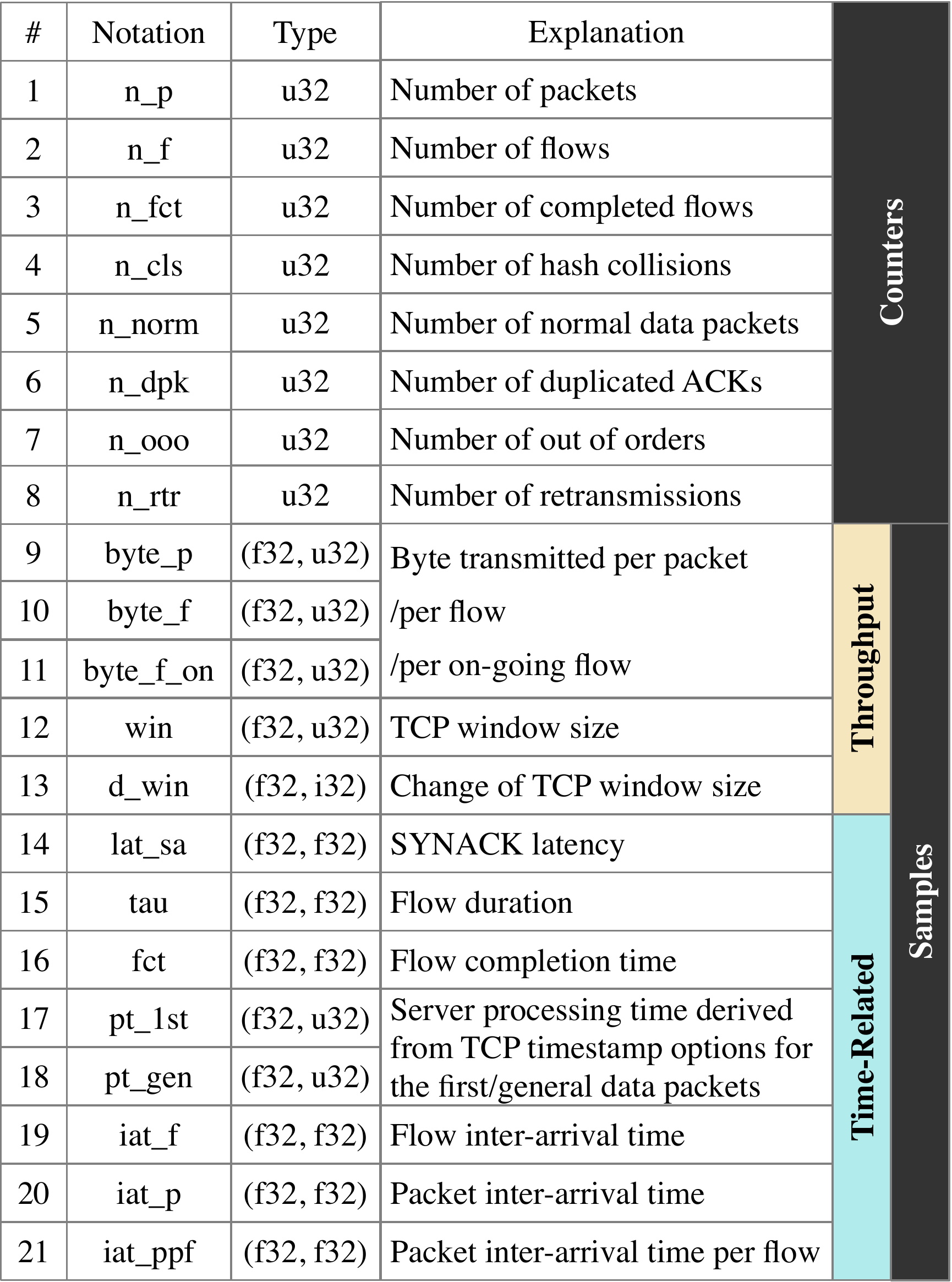}
    \caption{Feature list.}
		\label{fig:app-feature}
  \end{minipage}%
  \begin{minipage}{.5\textwidth}
    \begin{minipage}{\textwidth}
      \centering
      \includegraphics[height=1in]{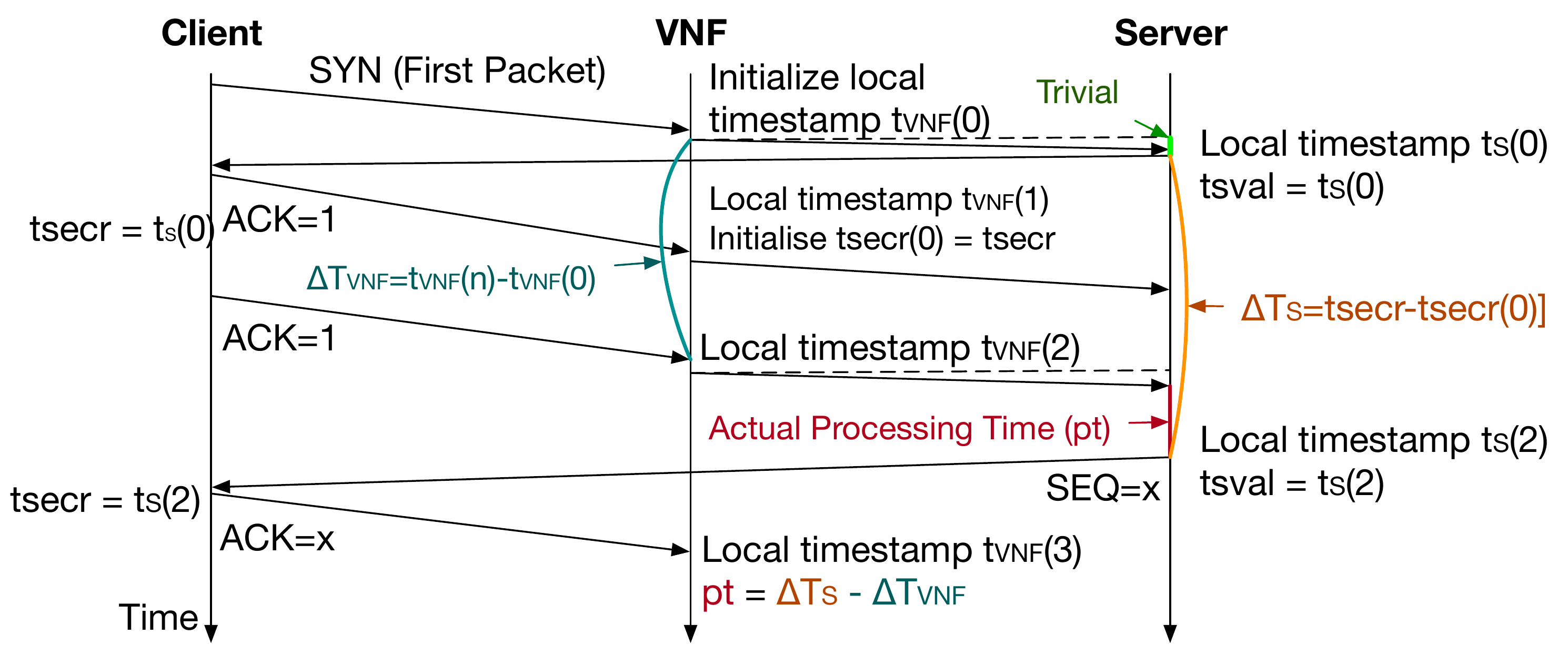}
      \caption{Processing time with TCP timestamp option.}
      \label{fig:app-pt}  
    \end{minipage}
    \begin{minipage}{\textwidth}
      \centering
      \includegraphics[height=1.5in]{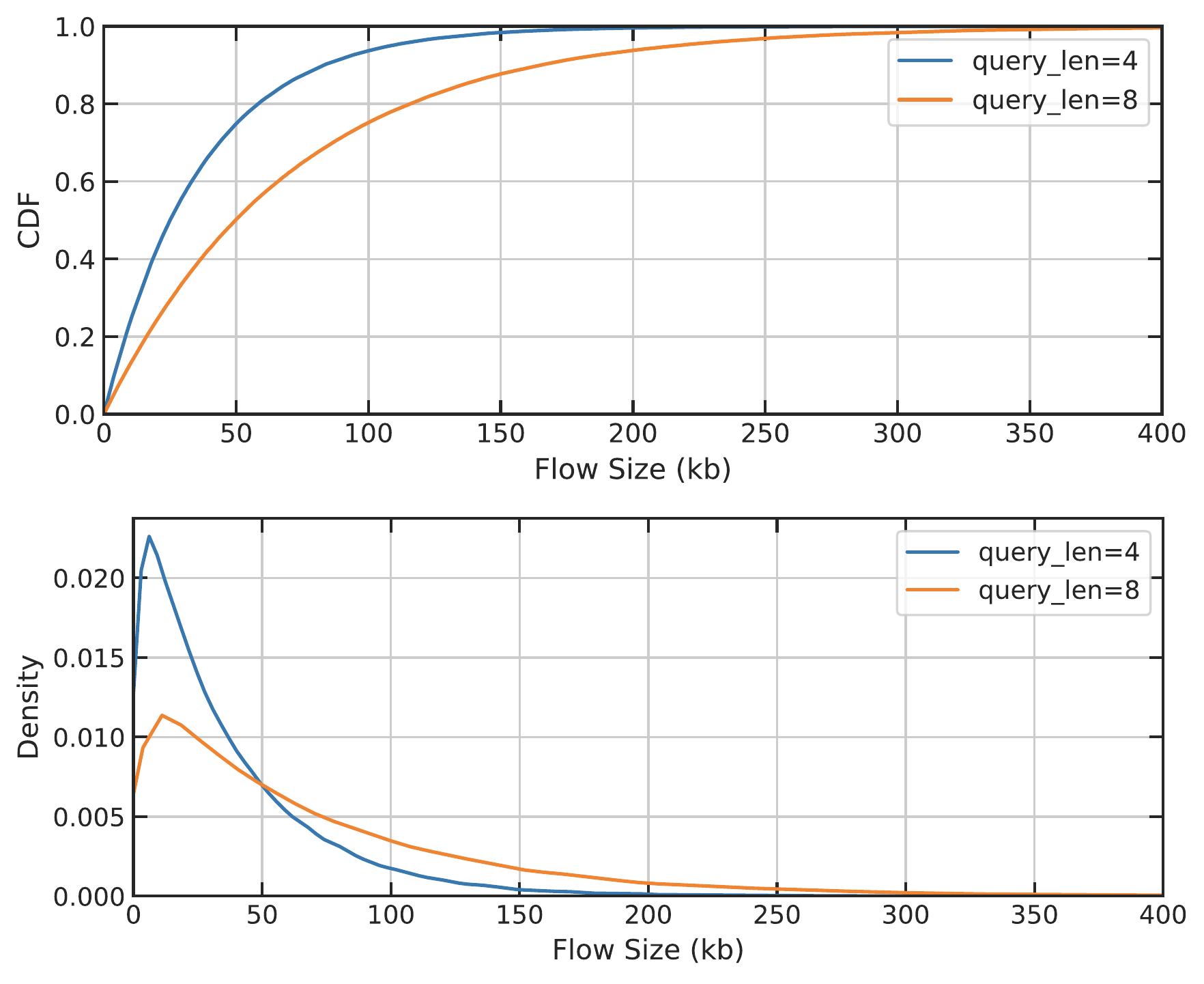}
      \caption{Flow size distribution.}
      \label{fig:app-flow-size}
    \end{minipage}  
  \end{minipage}
  \vskip -.1in
\end{figure}

The whole list of networking features coleected in \Albatross are shown in figure \ref{fig:app-feature}.
They can be categorized into counters and sampled features, which can be further divided into time- and throughput-related features.
Figure \ref{fig:app-pt} depicts the computation process of server processing time derived from TCP timestamp option.
TCP traffic is used in this paper to illustrate the workflow since it is the most widely used protocol in content delivery network~\cite{facebook-dc-traffic}.
The same workflow and discussion apply to other network traffic (UDP and QUIC).

\section{Testbed Configuration}
\label{app:testbed}

\subsection{System Platform}
\label{app:testbed-system}

Application servers are virtualized on $4$ UCS B200 M4 servers, each with one Intel Xeon E5-2690 v3 processor (12 physical cores and 48 logical cores), interconnected by UCS 6332 16UP fabric.
Operating systems are \texttt{Ubuntu 18.04.3 LTS} (\texttt{GNU/Linux 4.15.0-128-generic x86\_64}).
Compilers are \texttt{gcc} version \texttt{7.5.0} (\texttt{Ubuntu 7.5.0-3ubuntu1~18.04}).
Applications employed in this paper are the following: \texttt{Apache} \texttt{2.4.29}, \texttt{VPP v20.05}, \texttt{MySQL 5.7.25-0ubuntu0.18.04.2}, and \texttt{MediaWiki v1.30}.
The VMs are deployed on the same layer-$2$ link, with statically configured routing tables. 
Apache HTTP servers share the same VIP address on one end of GRE tunnels with the load balancer on the other end.

\subsection{Apache HTTP Servers}
\label{app:testbed-apache}

The Apache servers use \texttt{mpm\_prefork} module to boost performance.
Each server has maximum $32$ worker threads and TCP backlog is set to $128$.
In the Linux kernel, the \texttt{tcp\_abort\_on\_overflow} parameter is enabled, so that a TCP RST will be triggered when the queue capacity of TCP connection backlog is exceeded, instead of silently dropping the packet and waiting for a SYN retransmit.
With this configuration, the FCT measures application response delays rather than potential TCP SYN retransmit delays.
Two metrics are gathered as ground truth server load state on the servers: CPU utilization and instant number of Apache busy threads.
CPU utilization is calculated as the ratio of non-idle cpu time to total cpu time measured from the file \texttt{/proc/stat} and the number of Apache busy threads is assessed via Apache's \textit{scoreboard} shared memory.

\subsection{$24$-Hour Wikipedia Replay Trace}
\label{app:testbed-wiki}

To create Wikipedia server replicas, an instance of MediaWiki\footnote{https://www.mediawiki.org/wiki/Download} of version $1.30$, a MySQL server and the \texttt{memcached} cache daemon are installed on each of the application server instance. 
\textit{WikiLoader} tool~\cite{wikiloader} and a copy of the English version of Wikipedia database~\cite{wiki_traces}, are used to populate MySQL databases. 
The 24-hour trace is obtained from the authors of~\cite{wiki_traces} and for privacy reasons, the trace does not contain any information that exposes user identities.

\subsection{PHP \texttt{for}-Loop Trace}
\label{app:testbed-for-loop}

To study CPU-bound applications, a PHP \texttt{for}-loop script is used, whose requested number of iterations \texttt{\#iter} follows an exponential distribution.
The sizes of the queries' replies are proportional to the number of iterations.
This allows to generate a heavy-tail distribution of flow durations and transmitted bytes as in~\cite{facebook-dc-traffic}.

\subsection{PHP File Trace}
\label{app:testbed-file}

To simulate IO-bound applications, PHP queries for static files of different sizes are used as in~\cite{lbas-2020}.
The sizes of files are $100$KB, $200$KB, $500$KB, $750$KB, $1$MB, $2$MB, and $5$MB. 
$50$ files are generated for each size.

\subsection{Configurations for $3$ Traces}
\label{app:trace-config}

\begin{table}[t]
  \caption{Three configurations with different traces.}
  \label{tab:app-trace}
  \centering
  \begin{tabular}{cccc}
    \toprule
    \multicolumn{1}{c}{\begin{tabular}[c]{@{}c@{}}Trace\end{tabular}} & \multicolumn{1}{c}{\texttt{Wiki}} & \multicolumn{1}{c}{\texttt{for}-loop} & \multicolumn{1}{c}{file} \\ \midrule
    \multicolumn{1}{c}{Group $0$ servers}                                                   & \multicolumn{1}{c}{$3$ $\times$ $4$-CPU}  & \multicolumn{1}{c}{$36$ $\times$ $2$-CPU}  & \multicolumn{1}{c}{$36$ $\times$ $2$-CPU}  \\ 
    \multicolumn{1}{c}{Group $1$ servers}                                                   & \multicolumn{1}{c}{$4$ $\times$ $2$-CPU}  & \multicolumn{1}{c}{$24$ $\times$ $4$-CPU}  & \multicolumn{1}{c}{$24$ $\times$ $4$-CPU}  \\ 
      \multicolumn{1}{c}{Queries/s}                                                   & \multicolumn{1}{c}{$\left[369, 518\right]$}  & \multicolumn{1}{c}{$\left[350, 500\right]$}  & \multicolumn{1}{c}{$\left[400, 1000\right]$}  \\ \bottomrule
  \end{tabular}
  \vskip -0.2in
\end{table}

The corresponding configurations for the $3$ types of network traffic are listed in table \ref{tab:app-trace}.
The flow size distributions of $2$ types of Poisson \texttt{for}-loop traces are depicted in figure \ref{fig:app-flow-size}.

\section{Reproducibility}
\label{app:reproduce}

The proposed mechanism (\Albatrosss) and artifacts (including code and datasets) will be open-sourced at https://github.com/ZhiyuanYaoJ/Aquarius.
It allows to reproduce results in the submitted paper, and to generating more benchmark datasets that potentially benefit interdisciplinary research on computer networking and machine learning.
A pipeline will be provided and documented to make the best use of \Albatross for both offline and online applications.
Examples are illustrated by way of jupyter notebooks, which serve as step-by-step tutorials.

\section{ML Models}
\label{app:model}

Dense1 is a benchmark model for sequential with $1$ flatten layer and $1$ fully connected layer of neural network.
RNN2 is a model $2$ $20$-hidden-unit \texttt{SimpleRNN} layers (first layer with \texttt{return\_sequence=True}) and $2$ fully connected layers (the first layer with $32$ hidden units and the second as output layer).
LSTM2 replaces the \texttt{SimpleRNN} RNN2 model with LSTM layers and GRU2 with GRU layers.
1DConv-GRU1 applies 1-dimentional CNN (as in textCNN) before applying $1$ layer of GRU networks and $1$ fully connected layers (output layer).
1DConv-GRU1 is modified by adding $1$ GRU layer between the CNN and the GRU layer with \texttt{return\_sequence=True}, which gives limited improvement on sacrifying computational overhead.
This variant model is omitted in this paper.
Wavenet-GRU1 stacks $3$ dilated 1D convolutional layers with $1$ layer of $20$-hidden-unit GRU and $1$ fully connected layers (output layer).
Wavenet-Reconst puts $1$ $16$-hidden-unit embedding layer before $4$ stacked dilated 1D convolutional layers and $1$ layer of $20$-hidden-unit GRU layer, together with $1$ output layer.

\end{document}